\documentclass[journal=jpcl,manuscript=article]{achemso}
\usepackage[version=3]{mhchem} % Formula subscripts using \ce{}
\usepackage{color}
\usepackage{hyperref}
\usepackage{soul}
\usepackage{booktabs}

\author{Priyanka Arvind Paunikar}
\affiliation[HFML-FELIX]
{HFML-FELIX, Toernooiveld 7, 6525 ED Nijmegen, The Netherlands}
\alsoaffiliation[IMM]
{Institute for Molecules and Materials, Radboud University, Heyendaalseweg 135, 6525 AJ Nijmegen, The Netherlands}

\author{Hugo Maurer}
\affiliation[Van]
{Van 't Hoff Institute for Molecular Sciences, University of Amsterdam, Science Park 904, 1098 XH Amsterdam, The Netherlands}

\author{Lars Reems}
\affiliation[Amsterdam]
{Anton Pannekoek Institute for Astronomy, University of Amsterdam, Science Park 904, 1098XH Amsterdam, The Netherlands}

\author{Alessandra Candian}
\affiliation[Amsterdam]
{Anton Pannekoek Institute for Astronomy, University of Amsterdam, Science Park 904, 1098XH Amsterdam, The Netherlands}

\author{Sandra Brünken}
\affiliation[HFML-FELIX]
{HFML-FELIX, Toernooiveld 7, 6525 ED Nijmegen, The Netherlands}
\alsoaffiliation[IMM]
{Institute for Molecules and Materials, Radboud University, Heyendaalseweg 135, 6525 AJ Nijmegen, The Netherlands}

\author{Jos Oomens}
\affiliation[HFML-FELIX]
{HFML-FELIX, Toernooiveld 7, 6525 ED Nijmegen, The Netherlands}
\alsoaffiliation[IMM]
{Institute for Molecules and Materials, Radboud University, Heyendaalseweg 135, 6525 AJ Nijmegen, The Netherlands}

\author{Piero Ferrari}
\affiliation[HFML-FELIX]
{HFML-FELIX, Toernooiveld 7, 6525 ED Nijmegen, The Netherlands}
\email{piero.ferrariramirez@ru.nl}

\author{Wybren Jan Buma}
\affiliation[Van]
{Van 't Hoff Institute for Molecular Sciences, University of Amsterdam, Science Park 904, 1098 XH Amsterdam, The Netherlands}
\alsoaffiliation[IMM]
{Institute for Molecules and Materials, Radboud University, Heyendaalseweg 135, 6525 AJ Nijmegen, The Netherlands}
\email{w.j.buma@uva.nl}

\title[]
  {Interstellar Aromatic Infrared Bands: IR absorption spectroscopy of the lowest triplet state of naphthalene}

\begin{document}

%%%%%%%%%%%%%%%%%%%%%%%%%%%%%%%%%%%%%%%%%%%%%%%%%%%%%%%%%%%%%%%%%%%%%
\begin{abstract}
The aromatic infrared bands (AIBs), emission features observed ubiquitously across a wide variety of interstellar objects, are commonly attributed to radiative cooling of highly-excited ground-state vibrational levels of polycyclic aromatic hydrocarbons (PAHs) that are populated after UV absorption and internal conversion. However, for neutral PAHs intersystem crossing to the triplet manifold, followed by radiative cooling of vibrational levels within the lowest excited triplet state is expected to be a competing relaxation channel. This state differs in vibrational frequencies, transition moments and internal energy, and therefore gives rise to different emission spectra. Despite their significance, neutral PAHs in their triplet states have remained largely unexplored, primarily because of the lack of direct spectroscopic access. Here, we report the first measurement of the infrared absorption spectrum of the lowest triplet state of naphthalene. The spectrum is markedly different from the vibrational spectrum of the singlet electronic ground state, displaying features that can help identifying PAHs in the triplet states in astronomical spectra. We also show that experimental spectra are closely reproduced by anharmonic frequency calculations. These findings serve as a stepping stone for detecting triplet-state PAHs in the interstellar medium and provide a basis for developing more accurate infrared emission models. 
\end{abstract}

%%%%%%%%%%%%%%%%%%%%%%%%%%%%%%%%%%%%%%%%%%%%%%%%%%%%%%%%%%%%%%%%%%%%%
\section{Introduction}\label{sec1}
The interstellar medium (ISM) is chemically diverse, and this complexity is central to the formation and evolution of stars and planets. This chemical complexity is reflected in the ISM’s characteristic aromatic infrared bands (AIBs).\cite{allamandola1989interstellar} The general consensus is that these features originate from infrared emission of UV-pumped polycyclic aromatic hydrocarbons (PAHs),\cite{leger1984identification}\cite{tielens2026aromatic}\cite{tielens2008interstellar}\cite{bierbaum2011pahs} which may constitute up to 20\% of the total carbon budget in the ISM.\cite{joblin2009interstellar} As yet, it has been assumed that these emissions originate from radiative transitions between vibrational levels of the electronic ground state of PAHs, be it in their neutral or charged state. Neutral PAHs contribute significantly to the 3.3 $\mu$m (C-H stretching) and 11-15 $\mu$m (C-H out-of-plane bending modes) region, while cationic PAHs dominate the 6-9 $\mu$m (C-C stretching and C-H in-plane modes) region.\cite{bakes2001theoretical}\cite{bakes2001theoretica2l} Importantly, these mid-IR emission features are not molecule-specific, but rather functional-group-specific. Identification of the specific class of PAHs responsible for these AIBs therefore remains a challenging problem and a vibrant area of investigation in astrochemistry.\cite{tielens2026aromatic}\cite{sundararajan2025infrared}\cite{peeters2002rich}\cite{bouwman2021mid}\cite{khan2025pdrs4all}\cite{buragohain2018interstellar} So far, no individual PAH has been identified in the infrared, although several cyano-substituted PAHs have been identified via their rotational transitions.\cite{wenzel2024detection}

\begin{figure}
    \centering
    \includegraphics[width=0.7\linewidth]{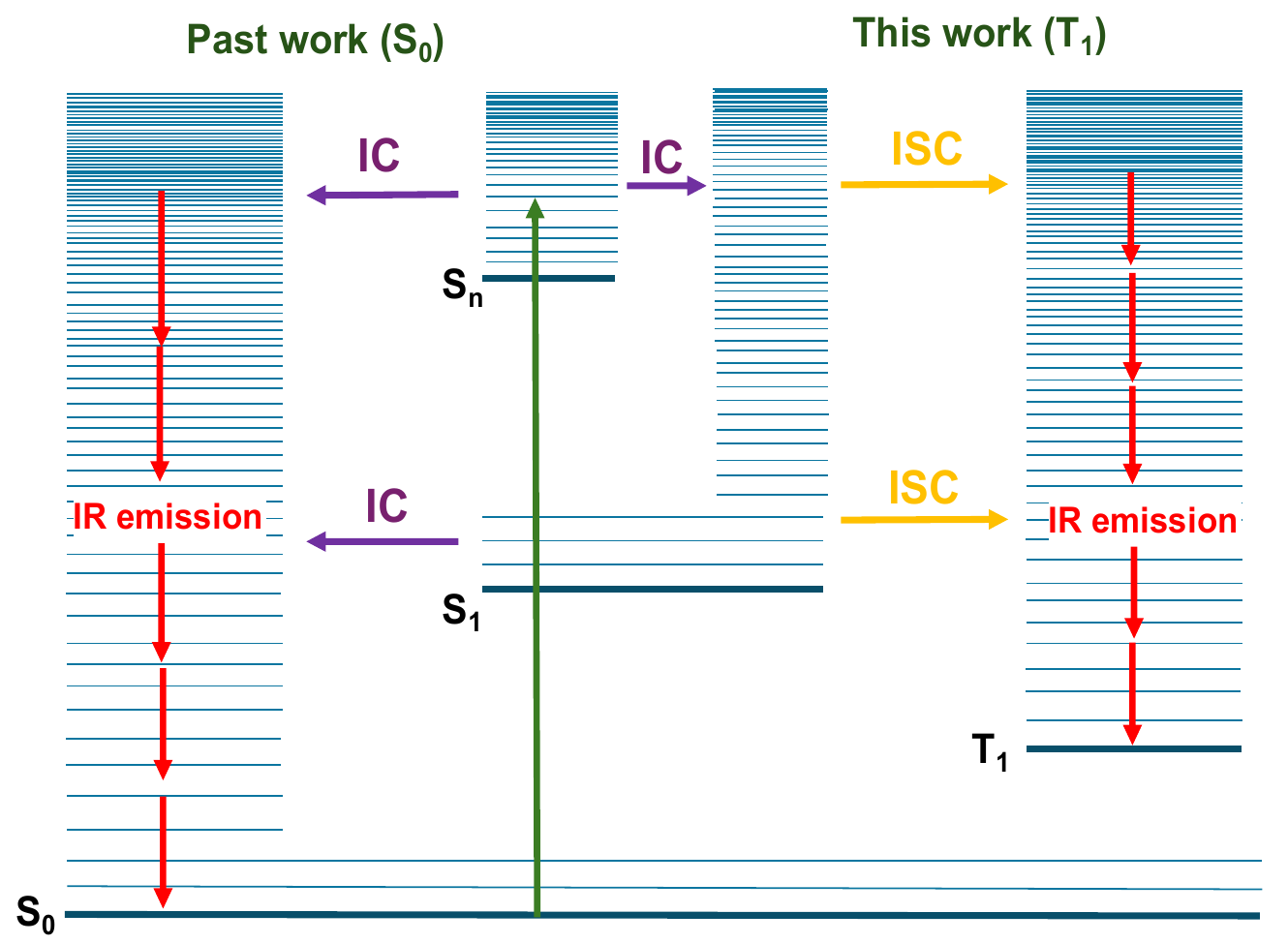}
    \caption{Jablonski diagram depicting the radiative and non-radiative relaxation channels of neutral polycyclic aromatic hydrocarbons (PAHs) following the absorption of a UV photon. In the figure, internal conversion (IC) is represented by a  purple arrow, while intersystem crossing (ISC) is depicted by a yellow arrow.}
    \label{Figure:1}
\end{figure}

Here, we focus on neutral PAHs, which are closed-shell species with a singlet electronic ground state, S$_{0}$ (Figure 1). Upon absorption of a UV photon (green arrow), a neutral PAH is excited to higher electronically excited singlet states, S$_{\textnormal{n}}$, from which rapid internal conversion (IC, purple arrows) occurs to lower-lying (vibrationally excited) singlet states (e.g. S$_{1}$) and ultimately to S$_{0}$. AIBs are attributed to radiative decay from hot vibrational levels of S$_{0}$ populated after internal conversion from S$_{\textnormal{n}}$ to S$_{0}$. However, a competing -and in many cases dominant\cite{dabestani1999compilation}- decay pathway from S$_{1}$ is intersystem crossing (ISC, yellow arrow) to the triplet manifold, leading to a population of highly-excited vibrational levels of the lowest excited triplet state T$_{1}$.\cite{zahlan1965singlet}\cite{ruiz2013singlet}\cite{johnson2015photo} (Non-)radiative decay rates of T$_{1}$ to S$_{0}$ are in general much slower than radiative rates within the T$_{1}$ vibrational manifold.\cite{tielens2026aromatic} A subpopulation of PAHs trapped in long-lived T$_{1}$ states would thus have major consequences for the interpretation of astronomical observations of AIBs, primarily because of the intrinsically different vibrational spectroscopy of T$_{1}$ and S$_{0}$ (i.e., vibrational frequencies, transition dipole moments, and anharmonicities). In addition, the difference in energy between T$_{1}$ and S$_{0}$ would affect the appearance of their emission spectra: since T$_{1}$ lies higher in energy than S$_{0}$, the amount of energy that can be released through radiative cooling is smaller, resulting in a lower "effective temperature" for the emitting PAHs. This affects band shapes and relative intensities.\cite{mackie2018anharmonic}

PAHs in triplet states can play an important role in the ISM. Since singlet and triplet states differ in their orbital occupancies and charge density distributions, neutral PAHs in different spin states may exhibit distinct chemical reactivities. However, the reactivity of triplet neutral PAHs under ISM conditions remains largely unexplored. Evidence from related systems, such as dehydrogenated PAH cations under ISM conditions, has shown that singlet and triplet states can exhibit substantially different reactivities toward H$_{2}$.\cite{snow1998interstellar}\cite{snow2008ion} In addition, computational studies of PAH growth involving neutral PAHs predict that triplet PAHs can react with unsaturated hydrocarbons, such as acetylene, through carbon-addition–acetylene-migration reactions.\cite{zhang2015role} Together, these results suggest that the spin state may play an important role in determining PAH reactivity and growth pathways in the ISM. 

It is also interesting to consider the potential role of triplet-state PAHs as carriers of Extended Red Emission (ERE). ERE is a broad red emission feature commonly observed in UV-irradiated regions such as photodissociation regions, and is generally attributed to excitation by far-ultraviolet photons. Proposed ERE carriers absorb photons in the 540–91.2 nm range and emit in the 600–850 nm range.\cite{lai2017extended} PAHs can absorb UV photons and undergo intersystem crossing (ISC) to long-lived, low-lying triplet states, making them plausible candidates in this context.

An investigation of the vibrational spectra of PAHs in their lowest excited triplet state is thus deemed essential for accurately interpreting and understanding AIBs, and indirectly for elucidating the inventory and evolution of carbon species in space. Earlier matrix‐isolation studies in N$_{2}$ and Ar, as well as transient polarized resonance Raman measurements in solution, have identified several vibrational bands of the lowest triplet states of small PAHs, including naphthalene and anthracene. These studies show that vibrational frequencies of the T$_{1}$ state differ significantly from those of the S$_{0}$ state, especially in the 1000–1500 cm$^{-1}$ region, where most bands were assigned.\cite{kudoh1999matrix}\cite{nakata2000lowest}\cite{hoesterey1987triplet}\cite{krumschmidt1991triplet}\cite{kumakura2018matrix} However, these works neither provide a complete assignment of the T$_{1}$ spectrum nor fully reflect the intrinsic gas-phase behavior of these species. It is therefore important to measure such spectra under astronomically relevant conditions, i.e., as isolated species and at low temperatures. The availability of such laboratory-based spectra has become even more crucial with the launch of the James Webb Space Telescope (JWST), whose spectral and spatial resolution has revolutionized studies of AIBs.\cite{rigopoulou2024polycyclic}\cite{khan2025pdrs4all}\cite{chown2024pdrs4all}

For neutral ground-state PAHs, well-resolved IR absorption spectra under astronomically relevant conditions can be routinely recorded nowadays. To this end, Resonance Enhanced MultiPhoton Ionization (REMPI) in combination with IR-UV depletion spectroscopy has been employed\cite{lemmens2020polycyclic}\cite{lemmens2023wetting} using tabletop infrared lasers to study the 3 $\mu$m region, and infrared free-electron lasers such as FELIX for the mid- and far-infrared range. This has resulted in a wealth of experimental and computational studies,\cite{sundararajan2025infrared}\cite{remmers2000gas}\cite{lemmens2020far}\cite{mackie2015anharmonic}\cite{maltseva2015high}\cite{cockett1993vibronic}\cite{lemmens2019anharmonicity}\cite{chakraborty2014theory}\cite{ricca2012infrared}\cite{ricca2010far}\cite{lemmens2021infrared}\cite{mackie2015characterizing}\cite{esposito2024anharmonic}\cite{sehring2025reparameterized}\cite{bauschlicher2008infrared}\cite{ricca2010far}\cite{ricca2012infrared} which have led to the conclusion, amongst others, that IR absorption and emission spectra are strongly affected by vibrational anharmonicities.\cite{maltseva2015high} IR absorption spectra of neutral PAHs in their long-lived T$_{1}$ state recorded under similar conditions are still notoriously missing, because sensitive methods for monitoring vibrational level populations in the T$_{1}$ state have thus far been largely unsuccessful, unlike the successful application of IR-UV depletion spectroscopy in molecular beams for S$_{0}$.  

In this work, we present an IR-UV REMPI depletion scheme that allows us to record the IR absorption spectrum of isolated molecules in their long-lived triplet electronic state. We illustrate its applicability by recording the IR absorption spectrum of T$_{1}$ in the 3 $\mu$m region and the fingerprint region (6.6-20 $\mu$m) of naphthalene – the smallest representative of the PAH family that can be used as a benchmark. We show that these spectra are accurately reproduced by Density Functional Theory (DFT) anharmonic vibrational frequency calculations, in particular in the fingerprint region. Experiments and calculations thereby provide unique insight into the wavelength regions where distinct differences are expected to occur in IR emission bands of S$_{0}$ and T$_{1}$. As such, these spectra may guide searches for triplet IR emission bands in astronomical observations.

\section{Results and discussion}\label{sec2}
We present and discuss our results in three steps. We first examine the ISC dynamics of naphthalene following excitation to S$_{1}$ leading to the formation of a long-lived T$_{1}$ state. Then we record IR absorption spectra in the T$_{1}$ state using IR-UV depletion spectroscopy with a tabletop OPO/OPA laser and the FELIX free-electron laser, and compare them with spectra predicted by anharmonic calculations of the vibrational spectra. (See the Methods section for details about the Experimental and Theoretical methods.) Finally, we discuss the astrochemical implications of the significantly different IR spectra in the S$_{0}$ and T$_{1}$ states.

\subsection{Electronic spectroscopy}\label{subsec2}
Figure A1  shows the onset of the S$_{1}$($^{1}$B$_{3u}$) $\leftarrow$ S$_{0}$($^{1}$A$_{g}$) two-color (1+1’) R2PI excitation spectrum, which reproduces previously reported spectra using LIF and REMPI spectroscopy.\cite{krumschmidt1991triplet}\cite{maltseva2015high}\cite{cockett1993vibronic} The same Figure also shows in panel (b) the ion signal as a function of the time delay between excitation and ionization upon excitation at the 0-0 transition at 32020 cm$^{-1}$. The decay of this signal shows two components, an exponentially decaying component with a lifetime of 0.27$\pm$0.02 $\mu$s and a constant signal that is limited by the measurement window of the instrument. The first component corresponds to the lifetime of the vibrationless S$_{1}$ state of naphthalene\cite{behlen1981intersystem}, whereas the second component is associated with ionization of molecules that have crossed to the T$_{1}$ state. This result is consistent with the findings of Johnson \textit{et al}.\cite{johnson2015photo} Based on this time profile, we find that time delays between excitation and ionization larger than 1 $\mu$s are suitable to perform spectroscopic measurements on naphthalene exclusively in the T$_{1}$ state.

\begin{figure}
    \centering
    \includegraphics[width=0.6\linewidth]{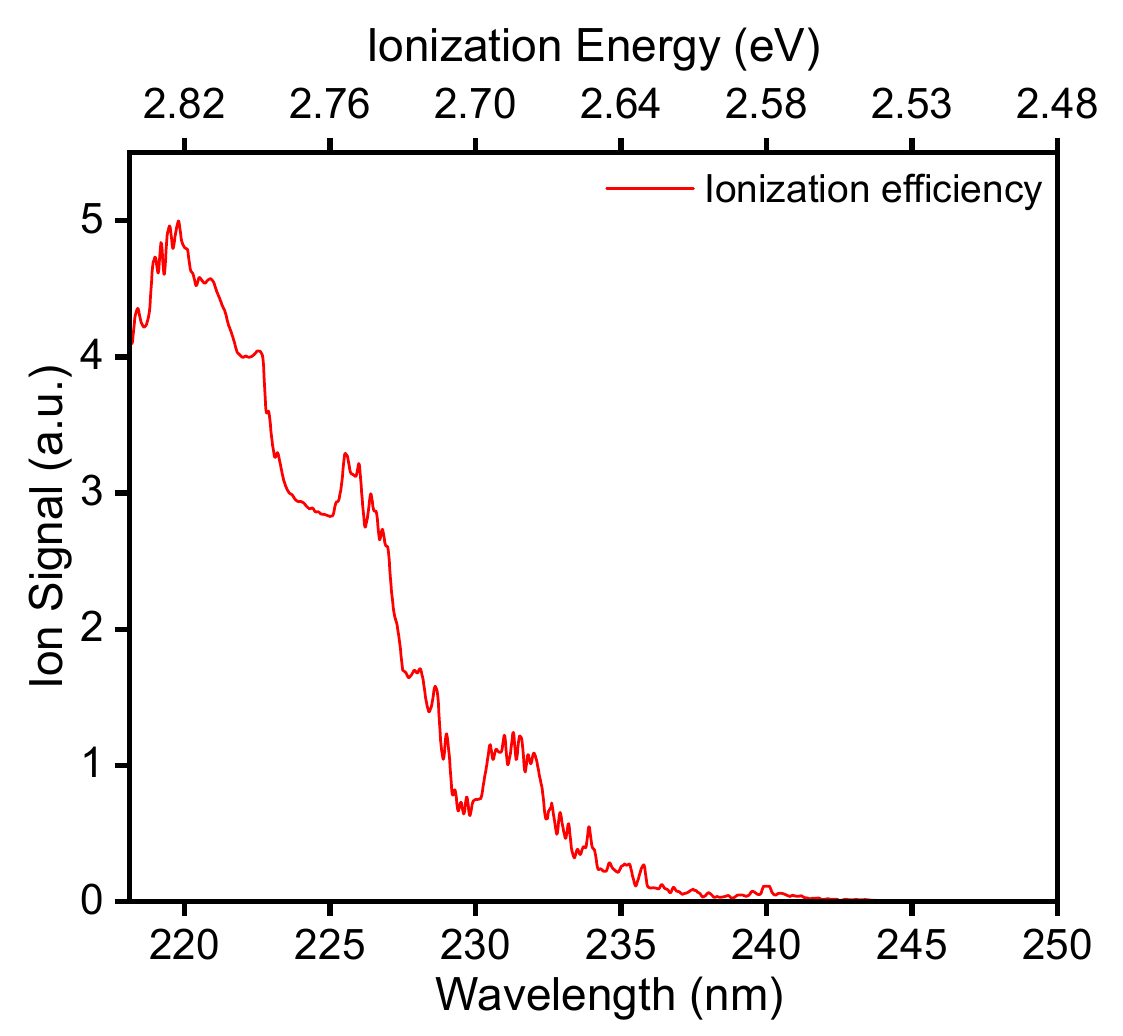}
    \caption{Ionization efficiency curve of the T$_{1}$ state of naphthalene in the 215-245 nm region.}
    \label{Figure:2}
\end{figure}

In Figure 2, we report the ionization efficiency curve of the T$_{1}$ state, recorded after state-selective excitation to the vibrational ground state of S$_{1}$ and delaying the ionization laser by 1 $\mu$s. This spectrum shows a number of bands associated with T$_{n}$ $\leftarrow$ T$_{1}$ absorptions that are broadly consistent with the previously reported data in this energy region.\cite{johnson2015photo} Before discussing the infrared experiments in detail, it is useful to mention a number of aspects. Following excitation to the S$_{1}$ state and under collisionless molecular beam conditions, the absorbed energy is retained upon radiationless internal conversion to the ground state (S$_{0}$) or ISC to the triplet manifold. ISC thus leads to a population of T$_{1}$ with an internal energy corresponding to the energy gap between S$_{1}$ and T$_{1}$ ($\sim$1.33 eV\cite{cockett1993vibronic}). Secondly, the adiabatic ionization energy from S$_{0}$ to the ground state of the cation D$_{0}$ is 8.144 eV (65687 cm$^{-1}$)\cite{cockett1993vibronic} implying that after excitation to S$_{1}$ (3.97 eV) and ISC to T$_{1}$ ionization could theoretically occur for excitation energies above 4.174 eV ($\sim$297 nm). However, on the basis of propensity rules for ionization\cite{blanchet2001electronic}\cite{schmitt2001electronic} that predict ‘conservation’ of internal energy upon ionization, ionization is expected to be most effective around 5.425 eV ($\sim$228 nm) explaining the strong wavelength dependence of the ionization cross section in the investigated wavelength region. 

\subsection{Infrared absorption spectroscopy of T$_1$}\label{subsec3}
To record infrared depletion spectra of T$_{1}$, one aims for a large change in the ionization cross section when vibrational levels are excited. In our experiments, we have therefore first determined the ionization wavelengths at which the \textit{relative depletion}, that is, the change in ion yield divided by the ion yield without IR excitation, is largest, and found this to be optimal in the 230-240 nm region. The gray trace in Figure 3(a) shows the infrared absorption spectrum of naphthalene in its T$_{1}$ state in the 3.33-3.18 $\mu$m (3000-3140 cm$^{-1}$) region associated with the CH-stretch vibrational modes. The spectrum has been obtained using 236 nm and 230 nm as the ionization wavelengths, and shows a maximum depletion of $\sim$30\% (Figure A2), indicating that the ionization energy does not alter the infrared spectral profile. To compare with the S$_{0}$ spectrum, the red trace displays the analogous spectrum of the ground electronic state as measured in the present experiments, which reproduces the spectrum reported previously.\cite{maltseva2015high} Although the lower concentration of molecules in the T$_{1}$ state reduces the signal-to-noise ratio compared to that in the spectrum of S$_{0}$, the spectrum is still of high quality, showing a large number of distinct bands whose positions are given in Table A1. Importantly and as expected, the T$_{1}$ spectrum is significantly different from the spectrum measured for S$_{0}$,  both in terms of band positions and intensity distribution. 

\begin{figure}
    \centering
    \includegraphics[width=0.7\linewidth]{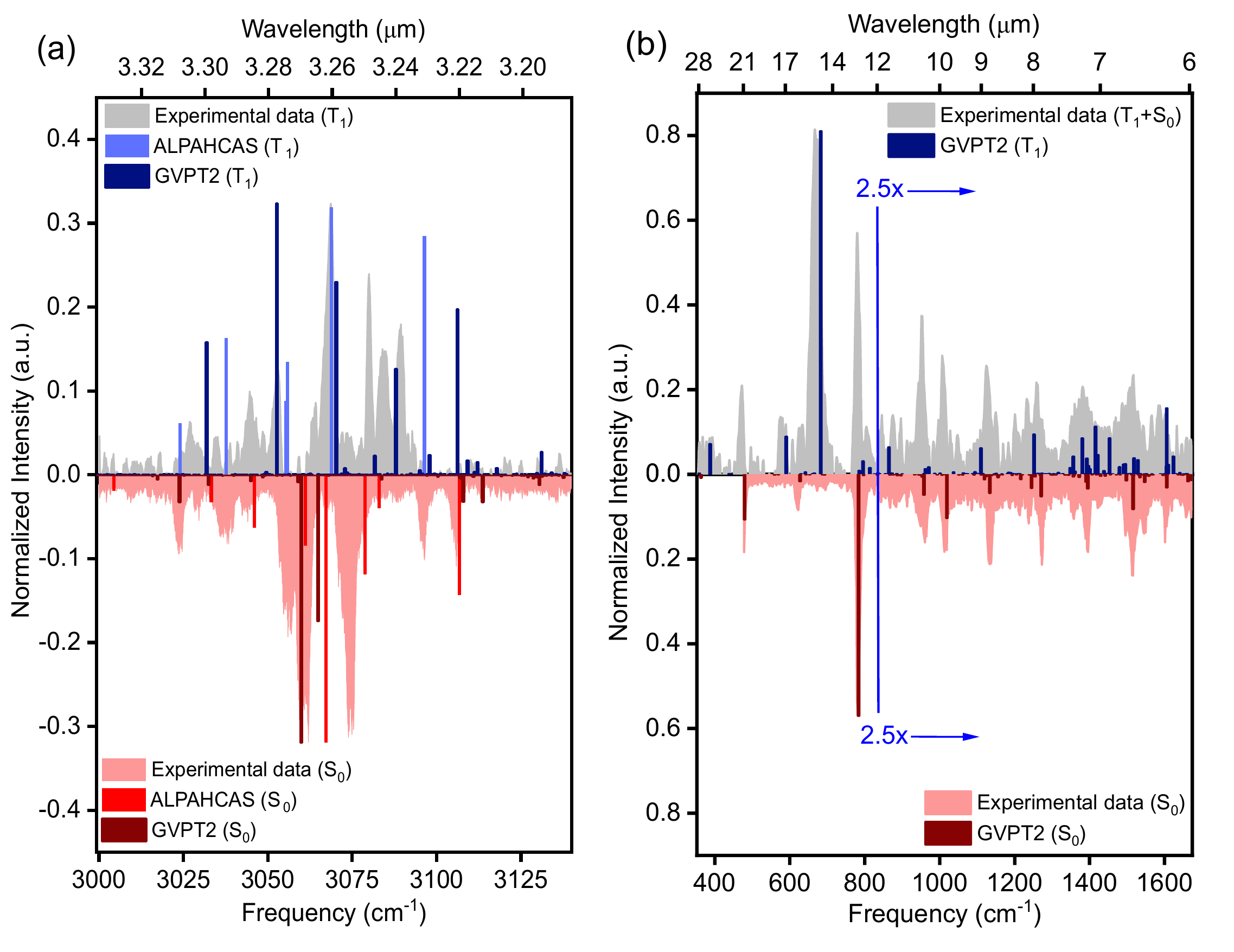}
    \caption{Experimental infrared absorption spectra of naphthalene in the T$_{1}$ (gray) and S$_{0}$ (red) states are shown in (a) the 3.33-3.18 $\mu$m (3000-3140 cm$^{-1}$) region and (b) the 28.6-5.88 $\mu$m (350-1700 cm$^{-1}$) region. In panel (a), the measured spectra are compared with anharmonic infrared spectra calculated using GVPT2 at the B3LYP/N07D level (dark blue and maroon stick spectra for T$_{1}$ and S$_{0}$, respectively) and with ALPAHCAS calculations (Reems \textit{et al.}, in prep.), which explicitly account for intensity redistribution arising from resonances (blue and red for T$_{1}$ and S$_{0}$, respectively). In panel (b), the measured spectra are compared only with the GVPT2 calculations. For clarity, the relative intensities of the bands in the 830-1674 cm$^{-1}$ region are scaled by a factor of 2.5.}
    \label{Figure:3}
\end{figure}

Anharmonic vibrational spectra predictions in S$_{0}$ and T$_{1}$ support these differences, as seen from the stick spectra in Figure 3(a), which are calculated using two different methodologies: GVPT2 and ALPAHCAS (VPT2 with polyads). (See the Methods section for further clarification of the differences between these two methods, and Figure A4 in the SI for the equilibrium structures and structural parameters of naphthalene in the two states.) Figure A7 shows the band assignments (labels) for the T$_1$ and S$_0$ spectra. The corresponding tabulated data, including wavenumbers and intensities, and other spectral parameters are provided in Table A1(a) for S$_0$ and Table A1(b) for T$_1$. In previous IR absorption studies on S$_{0}$, it was concluded that this region of the spectrum is strongly affected by anharmonic couplings, leading to Fermi resonances between fundamental modes with double as well as triple combination modes.\cite{maltseva2015high}\cite{mackie2015anharmonic} The same conclusion can now be drawn for the T$_{1}$ spectrum. Scrutinizing the predicted spectra, we find that inclusion of three-quanta combinations generates additional bands for both the S$_0$ and T$_1$ states. For S$_0$, a band appears at 3113.7 cm$^{-1}$, as previously anticipated in studies where three-quanta combination bands could not yet be explicitly included.\cite{mackie2015anharmonic,maltseva2015high} However, this predicted feature does not correspond to any of the experimentally observed bands. In contrast, for T$_1$, the inclusion of three-quanta modes gives rise to bands at 3081.7 and 3087.8 cm$^{-1}$, both of which coincide with experimentally observed spectral features. Thus, while three-quanta calculations predict additional bands for both electronic states, only those predicted for T$_1$ improve the agreement with experiment.

Many of the bands observed in 3 $\mu$m region of the S$_{0}$ and T$_{1}$ IR spectra can be reasonably well assigned by the calculations. The calculated spectra reproduce the observed band positions with a mean absolute deviation of $\sim$2-4  cm$^{-1}$ for the S$_{0}$ state and $\sim$1-6 cm$^{-1}$ for the T$_{1}$ state. Although GVPT2 and ALPAHCAS reproduce many common bands, there remain several experimental features that cannot be assigned (see Figure A7 and Table A1(a) and (b)). We note that the S$_{0}$ spectra for several larger PAHs show better agreement with theory.\cite{mackie2015anharmonic}\cite{maltseva2015high}\cite{lemmens2019anharmonicity}\cite{mackie2016anharmonic}\cite{mackie2018anharmonic}\cite{maltseva2016high}\cite{maltseva2018high} As such, and to benchmark the quality of predicted T$_{1}$ spectra in the 3 $\mu$m region, experiments on triplet states of other PAHs are much needed, and are now made feasible by the method presented here. 

Figure 3(b) shows the IR spectrum of T$_{1}$ (gray) and S$_{0}$ (red) of naphthalene in the 28.6-5.88 $\mu$m (350-1700 cm$^{-1}$) region. Due to the configuration of the laser systems used for the experiments with FELIX (see Methods section), the spectrum reported for T$_{1}$ actually contains bands from both the S$_{0}$ and T$_{1}$ states. Comparison of this spectrum with the S$_{0}$ spectrum enables us, however, to decompose this spectrum and identify the contribution of the two states to the various bands. The calculated spectra reproduce all observed band positions with a mean absolute deviation of $\sim$3 cm$^{-1}$ for the S$_{0}$ state and $\sim$10 cm$^{-1}$ for the T$_{1}$ state. (see Figure A8 and Table A2(a) and (b)). Such a decomposition rapidly leads to the conclusion that the pronounced band at 667 cm$^{-1}$ and the somewhat weaker bands at 390 and 578 cm$^{-1}$ are uniquely associated with T$_{1}$, and would thus provide unique identification markers for the presence of naphthalene in the T$_{1}$ state. In the 800–1600 cm$^{-1}$ range, there are many bands with much lower intensities. Due to the spectral bandwidth of FELIX (0.5-1\%, thus increasing at higher frequencies), an unambiguous identification and assignment of these bands to either T$_{1}$ or S$_{0}$ is more challenging. 

Previously, we have shown that anharmonic calculations of the S$_{0}$ IR absorption spectrum of naphthalene in this frequency region yield excellent agreement with the experimentally observed spectrum.\cite{lemmens2019anharmonicity} Similarly, we conclude that theory reproduces the experimental T$_{1}$ spectrum with the same accuracy. This is an important observation since, in general, calculations on electronically excited states are more prone to inaccuracies than calculations on electronic ground states. It also suggests that one can confidently make use of theoretically predicted spectra for PAHs for which experimental spectra may not be available yet.

\section{Astrophysical implications}\label{sec3}
Our results provide the first laboratory insights into the gas-phase infrared properties of the triplet states of polycyclic aromatic hydrocarbons (PAHs), which are crucial for identifying and accounting for triplet-state contributions in astronomical emission models. Although individual PAH species have not yet been identified in the ISM via their IR emission bands\cite{peeters2021spectroscopic} -due to the rather generic nature of their vibrational spectra and their near-zero permanent dipole moment making detection through their pure rotational spectrum impractical- our results highlight spectral bands unique to the triplet state of naphthalene. There are two key questions we aim to address here: (i) What is the role of PAHs in triplet states on the physical and chemical evolution of their astronomical environments (reactivity, carbon budget, stability, etc.)? (ii) More broadly, can PAHs in their triplet state be distinguished in AIB spectra, and are there specific spectral markers that could be used to identify them?

PAH emission models are useful tools for predicting emission profiles of PAHs across different internal energies or temperatures and, when combined with ionization balance calculations, can be used to constrain their charge distributions and abundances in various astrophysical environments.\cite{li2001infrared}\cite{andrews2016hydrogenation}\cite{montillaud2013evolution} Existing emission models often rely on a limited number of laboratory spectra and on harmonic-frequency calculations not accounting for anharmonicity. Moreover, these models generally ignore ISC and the resulting triplet-state emission. An exception is the study of  Falvo \textit{et al.}\cite{falvo2012probing} where both S$_{0}$- and T$_{1}$-state infrared emission spectra of small PAHs were modeled using calculated harmonic frequencies and a Monte Carlo approach to predict the anharmonic spectra. These calculations reveal that the emission spectra of the S$_{0}$ and T$_{1}$ states differ significantly, with most notable differences occurring in the C-H out-of-plane bending region, consistent with our experimental findings. Nevertheless, this study did not account for the large number of resonances, which are crucial for accurately capturing anharmonic effects, especially in the 3 $\mu$m region. Hence, laboratory data are key for benchmarking computational models and setting up an accurate formalism to estimate the stability and abundance of triplet PAHs in the ISM. 

The differences in the infrared absorption spectra of the singlet and triplet states of neutral PAHs arise from their distinct electronic properties. Their potential energy surfaces are different, and they therefore have different force constants and hence vibrational transitions. In addition, because of their different electronic distributions, also vibrational transition dipole moments -and thus vibrational transition intensities- are different. This is carried over into the infrared emission spectrum. The emission profile is also affected by internal energy, since the S$_{0}$ and T$_{1}$ states have different amounts of internal energy. The different internal temperatures of the two states lead to a different distribution of vibrational energy over the various modes. These properties are reflected in our measured infrared spectrum of the T$_{1}$ state of naphthalene, where significant differences are observed, for example, in the CH out-of-plane bending region around 667 cm$^{-1}$. This discussion can be extended to larger PAHs, as addressed below.

A key question that follows is whether triplet PAHs can be detected in the ISM. Their probability of detection depends on the UV absorption cross-section of the S$_{0}$ state, the efficiency of intersystem crossing from the S$_{1}$ state, the lifetime of the T$_{1}$ state, and the transition strength of the infrared modes. The S$_{1}$ $\leftarrow$ S$_{0}$ UV absorption cross-section determines the excitation rate, while the population of the triplet state is determined by the triplet quantum yield and its lifetime. The triplet quantum yield, which represents the efficiency of ISC to the T$_{1}$ state, is significant for many PAHs. Naphthalene -the smallest representative PAH- for example, has a triplet quantum yield of $\sim$0.75 in benzene solution\cite{dabestani1999compilation} while coronene, a larger PAH, has a triplet quantum yield of $\sim$0.61 in polymer matrices and slightly lower in the gas phase.\cite{o1980photomagnetism} The S$_{1}$–T$_{1}$ energy gap decreases with increasing PAH size, which enhances spin–orbit coupling between these states and facilitates efficient ISC.\cite{tielens2026aromatic} It is thus reasonable to expect that upon photon absorption, triplet-state PAHs are generated in a significant amount in the interstellar medium. PAHs in the triplet state are generally long-lived, typically on the order of seconds. For instance, the lifetime of the T$_{1}$ state of naphthalene and coronene is $\sim$2.6 s and $\sim$9.4 s, respectively.\cite{krumschmidt1991triplet}\cite{mcclure1949triplet} In solution, the triplet states are effectively quenched by collisions, but under the low-density conditions of the ISM, collisional quenching is virtually absent, and it is likely that the T$_{1}$ state emits radiatively in the IR or undergoes reverse intersystem crossing.\cite{tielens2026aromatic} Finally, the modeling study by Falvo \textit{et al.} estimated that the emission intensity of the C-H out-of-plane bending modes in the T$_{1}$ state modes can be 10–20\% higher than that of the corresponding S$_{0}$ vibrational transitions.\cite{falvo2012probing} This enhancement appears to arise from the larger intrinsic IR intensities in the T$_{1}$ state rather than from differences in the initial vibrational energy of the emission cascade. In our calculations, the strongest bands reach about 130 km mol$^{-1}$ in the T$_{1}$ state, compared with about 110 km mol$^{-1}$ in the S$_{0}$ state (Fig. A5), which is consistent with this interpretation.

The above argumentation raises the question of whether existing astronomical data may already contain signatures of triplet PAHs. We therefore computed vibrational spectra for six representative molecules in both their lowest singlet and triplet states (Fig. A6). In the triplet state, compared to the singlet, the IR spectra show the emergence and enhancement of two recurring diagnostic modes: the C–H out-of-plane bending mode (600–900 cm$^{-1}$; 16.7–11.1 $\mu$m) and a combined C–C stretching and C–H in-plane bending mode (1100–1500 cm$^{-1}$; 9.1–6.7 $\mu$m). For naphthalene, these appear at 667 cm$^{-1}$ (14.99 $\mu$m) and 1430 cm$^{-1}$ (6.99 $\mu$m), respectively. Although the intensity of the latter is relatively low, the calculations for larger PAHs predict that both bands remain present, with the intensity of the latter (1100–1500 cm$^{-1}$) increasing with PAH size. For neutral PAHs in their singlet state, the most prominent spectral features are typically found in the 11–15 $\mu$m region (C–H out-of-plane bending modes), while the 6–9 $\mu$m region (C–C stretching and C–H in-plane bending modes) shows comparatively weaker emission.\cite{khan2025pdrs4all} 

To assess the implications of these observations for the interpretation of existing astronomical data, we compared JWST/MIRI spectra of the Orion Bar, specifically the atomic PDR region, with averaged IR spectra of singlet and triplet PAHs, convolved with a Gaussian of 30 cm$^{-1}$ FWHM (see Fig. A9). We focus on the atomic PDR because it is directly exposed to stellar far-ultraviolet (FUV) radiation (6–13.6 eV), producing gas temperatures ranging from hundreds to a few thousand Kelvin at irradiated surfaces. These conditions provide the FUV field required to drive PAH excitation, ionization, and photochemical evolution, with excitation followed by efficient intersystem crossing (ISC) into triplet states, making the atomic PDR the most suitable environment for comparing laboratory studies of triplet PAHs with astronomical observations.\cite{peeters2024pdrs4all} The resulting comparison (Fig. A9) shows clear differences between singlet and triplet states, particularly for the triplet in the 1400–1600 cm$^{-1}$ region (7.1–6.2 $\mu$m), where we observe a broad feature centered at $\sim$6.8 $\mu$m (1480 cm$^{-1}$). This broad feature is comparable in intensity to those in the 11–15 $\mu$m range, whereas the singlet spectra do not exhibit comparable structure. 

The feature at $\sim$6.8 $\mu$m in the triplet-state calculations coincides with the small feature at 6.8 $\mu$m in the observational spectra. It also lies close to the observed well-known 6.2 $\mu$m band, which is commonly attributed to cationic PAHs and is therefore not present in the neutral singlet-state simulations. The intensity ratio of the 6.2 $\mu$m band to the 11.2 $\mu$m band is used to determine the ionization parameter, which reflects the balance between photoionization and recombination in the surrounding environment and is linked to the fraction of emission coming from cationic PAHs.\cite{maragkoudakis2026pdrs4all} Spectral decomposition tools used to assign PAH charge states include pyPAHdb, which fits observed spectra using a library of theoretically computed PAH spectra from the NASA Ames database, and template-based approaches such as PAHTAT, which derive templates from observations.\cite{ricca2026nasa}\cite{maragkoudakis2026pdrs4all}\cite{foschino2019learning} A common limitation is that neither one includes spectra of neutral PAHs in the triplet state, since no such database currently exists. Based on current knowledge, in the atomic PDR region, 60-70\% of the emission is from cations and 25\% is from neutrals.\cite{maragkoudakis2026pdrs4all} The assignment of the 6-9 $\mu$m cationic PAHs is based on the assumption that neutral PAHs contribute only weakly in the 6-9 $\mu$m region\cite{peeters2024pdrs4all}, an assumption derived exclusively from singlet state spectra. The present calculations show that triplet-neutral PAHs emit in the 6–9 $\mu$m region, with increasing intensity for larger molecules. This implies that current charge-state assignments in this spectral window might not be correct since a contribution from triplet neutral PAHs cannot be excluded. 

With the launch of JWST and the improved spectral resolution and sensitivity of the MIRI spectrograph (R $\approx$ 1500–3500\cite{nayak2024jwst}), it is now possible to better resolve PAH sub-features.\cite{khan2025pdrs4all}\cite{chown2024pdrs4all} The distinct feature at 6.8 $\mu$m in the calculated spectra, along with multiple bands in the 12–16 $\mu$m range, could also include contributions from triplet PAHs. Overall, it is plausible that triplet PAH emission features are already present in astronomical spectra but have remained unassigned, largely due to the lack of a comprehensive database of IR spectra for larger triplet PAHs,\cite{ricca2026nasa}\cite{malloci2007line} which will be essential for future identification. Further investigation of the presence of triplet PAH emission features requires dedicated modeling efforts.

To summarize, our results indicate that triplet states may play a more significant role in the photophysics of PAHs than has been assumed so far. The present results unambiguously demonstrate that the population of these states gives rise to infrared features that differ from those of the singlet ground state. In astrophysical environments, this raises the possibility that AIBs could well have contributions from vibrational emissions in T$_{1}$ instead of only S$_{0}$ as assumed up till now.  The presence of these states may also influence the survival of smaller PAHs under UV irradiation. For small PAHs such as naphthalene, where ISC is efficient and the T$_{1}$ lifetime is long, the triplet state represents a distinct reservoir of internal energy with potentially different dissociation pathways and timescales, motivating further investigation of its role in the photophysical evolution of small PAHs in space. More broadly, these findings point to a more nuanced picture of the evolution of carbonaceous molecules in space, in which intersystem crossing constitutes a relevant pathway. Incorporating such processes will be important for connecting molecular-scale photophysics to the large-scale evolution of carbon. 

\section{Conclusions}\label{sec3}
We have reported the first measurement of the IR absorption spectrum in the lowest-energy triplet electronic state of a prototypical PAH under astronomically relevant conditions. We have shown that the IR features of the T$_{1}$ state differ significantly from those of the S$_{0}$ state. In particular, the CH out-of-plane (600–900 cm$^{-1}$) and in-plane (1400–1500 cm$^{-1}$) bending regions provide electronic structure-dependent diagnostics. For example, the distinct band at 667 cm$^{-1}$ (14.99 $\mu$m) (CH out-of-plane mode) could serve as a potential marker for the astronomical detection of naphthalene in its excited triplet electronic state. The emission spectra of triplet PAHs mirror the differences in their absorption spectra, resulting in distinct band positions relative to the singlet ground state.\cite{falvo2012probing} Incorporating triplet-state PAHs into emission models will lead to revised abundances, e.g., singlet vs. triplet. Additionally, incorporating the triplet state into PAH photochemical models will influence the inferred chemistry of the interstellar medium, since PAHs behave differently in reactions and fragmentation between their singlet and triplet states. A more complete picture of triplet PAHs will require laboratory data on larger systems. This is particularly important for assessing how molecular size and structure (compactness) influence the AIBs. Work in this direction is already underway, with current efforts focused on extending measurements to larger PAHs. Taken together, these results suggest that existing models of PAH infrared emission need to be revisited to account for contributions from neutral triplet states. With the spectral resolution now available from the JWST, searching for signatures of triplet PAHs in observational data is a realistic next step. 

\section{Methods}\label{sec3}
\subsection{\textit{Experimental}}\label{subsec2}
The experiments were performed using a molecular beam spectrometer located at the HFML-FELIX institute in Nijmegen, The Netherlands.\cite{bakels2020gas} Naphthalene was heated to 80 $^{\circ}$C in an external reservoir and expanded into vacuum via a pulsed valve (Series 9 from General Valve) operating at 10 Hz using Ar at 4 bar as carrier gas. In this supersonic expansion molecules are adiabatically cooled down to rotational and vibrational temperatures of $\sim$5 and $\sim$20K, respectively, and are isolated, thus enabling spectroscopic studies under astronomically relevant conditions.\cite{segev2017molecular} After being collimated by a 2 mm skimmer, the molecular beam enters the ionization region of a reflectron time-of-flight mass spectrometer (R. M. Jordan D-850) where excitation and ionization takes place.

To probe the triplet state of neutral naphthalene, the sample is first excited to the S$_1$ state through its 0-0 transition at 321.30 nm (32020 cm$^{-1}$).\cite{cockett1993vibronic} This is done by exciting the molecular beam perpendicularly with the frequency-doubled output of a dye laser (LiopStar from Lioptec) operating on DCM in ethanol. The excited molecules undergo intersystem crossing to the T$_1$ state, located 1.33 eV below S$_1$\cite{cockett1993vibronic}\cite{chakraborty2014theory} where they are ionized using either the output of an ArF laser (193 nm) or the frequency-doubled output of a second dye laser operating on Coumarin 102 in ethanol.

IR absorption spectra of T$_{1}$ are obtained using IR-UV depletion spectroscopy by firing the IR laser in between the (S$_{0}$) excitation and (T$_{1}$) ionization lasers and monitoring the ion yield alternatingly with and without the IR laser. For the measurement of the IR spectrum in the 3000-3160 cm$^{-1}$ region, a tabletop OPO/OPA laser (LaserVision) was employed. This system delivers ns laser pulses with an average energy of 15 mJ in the scanned range with a spectral width of 0.1 cm$^{-1}$. The OPO/OPA system, the excitation and ionization lasers, and the molecular beam setup are operated at a repetition rate of 10 Hz. The IR spectrum is recorded by consecutively measuring mass spectra with and without IR irradiation by blocking every other infrared pulse with a mechanical shutter. In this case, the IR laser is aligned counterpropagating with respect to the excitation and ionization lasers and perpendicular to the molecular beam as depicted in Figure A3(a), and is delayed by 0.5 $\mu$s with respect to the excitation pulse.  

For the IR measurements in the 350-1700 cm$^{-1}$ region, FELIX is used as IR laser source. FELIX has a pulse structure composed of 10 $\mu$s macropulses with a spectral bandwidth of 0.5\% of the central wavelength and a typical energy of 60 mJ/pulse. Given the much longer macropulses of FELIX in comparison with the OPO/OPA laser, the optical alignment in these measurements is different. As shown in Figure A3(b), the FELIX beam is counterpropagating with respect to the molecular beam, and therefore perpendicular to the excitation and ionization lasers. In addition, the excitation laser is set to interact with the molecular beam before it enters the skimmer, thus ensuring a good temporal overlap between the excited molecules and the FELIX pulse. Although this means that a longer time delay between excitation and ionization needs to be used (310 $\mu$s), the long lifetime of T$_{1}$\cite{mcclure1949triplet} ensures that this increase in time delay does not affect the measurement. However, due to the configuration of the lasers and their temporal characteristics, it cannot be avoided that some fraction of the molecular beam that is probed is still in the S$_{0}$ state.    

\subsection{\textit{Theoretical}}\label{subsec2}
The anharmonic spectra presented in this paper are obtained via generalized second-order vibrational perturbation theory (GVPT2) applied to Density Functional Theory (DFT). Quartic Force Fields are generated using B3LYP/N07D\cite{barone2008development} and the default settings. Deviations from the default settings for Gaussian16 are "opt=verytight int(grid=200974) freq(HPmodes)”, and “Spectro=MaxQuanta=3” for the GVPT2 method. We used two different applications of this method: one implemented in Gaussian 16\cite{g16} and the other in the in-house code ALPAHCAS (Reems \textit{et al.}). In the Gaussian implementation, near-resonant terms (e.g., from Fermi and Darling Dennison resonances) involving fundamentals, overtone and combination bands up to three quanta are included in a small effective Hamiltonian matrix and diagonalized in that subspace. In ALPAHCAS, all resonant terms involving fundamentals and 2-quanta modes are identified and collected in polyads. These polyads represent the basis on which an effective Hamiltonian matrix is built and then diagonalized. This last treatment has been proven very effective in reproducing the highly anharmonic C-H stretch region\cite{mackie2018anharmonic}. For the computations with ALPAHCAS, a threshold of 200 cm$^{-1}$ is used for both the resonances and the polyads.

%%%%%%%%%%%%%%%%%%%%%%%%%%%%%%%%%%%%%%%%%%%%%%%%%%%%%%%%%%%%%%%%%%%%%
\begin{acknowledgement}
This work has been supported by the Nederlandse Organisatie voor Wetenschappelijk Onderzoek (NWO) Dutch Astrochemistry Network (grant no. ASTRO.JWST.001) and is based upon work from COST Action CA21126 - Carbon molecular nanostructures in space (NanoSpace), supported by COST (European Cooperation in Science and Technology). The authors gratefully acknowledge NWO for the support of the HFML-FELIX Institute and for CPU time on the Dutch National Supercomputer Snellius (NWO Rekentijd 2024.009), as well as the constant help of the HFML-FELIX staff.
\end{acknowledgement}
%%%%%%%%%%%%%%%%%%%%%%%%%%%%%%%%%%%%%%%%%%%%%%%%%%%%%%%%%%%%%%%%%%%%%

\section{Supplementary Information}\label{secA1}

\begin{figure}
    \centering
    \includegraphics[width=0.8\linewidth]{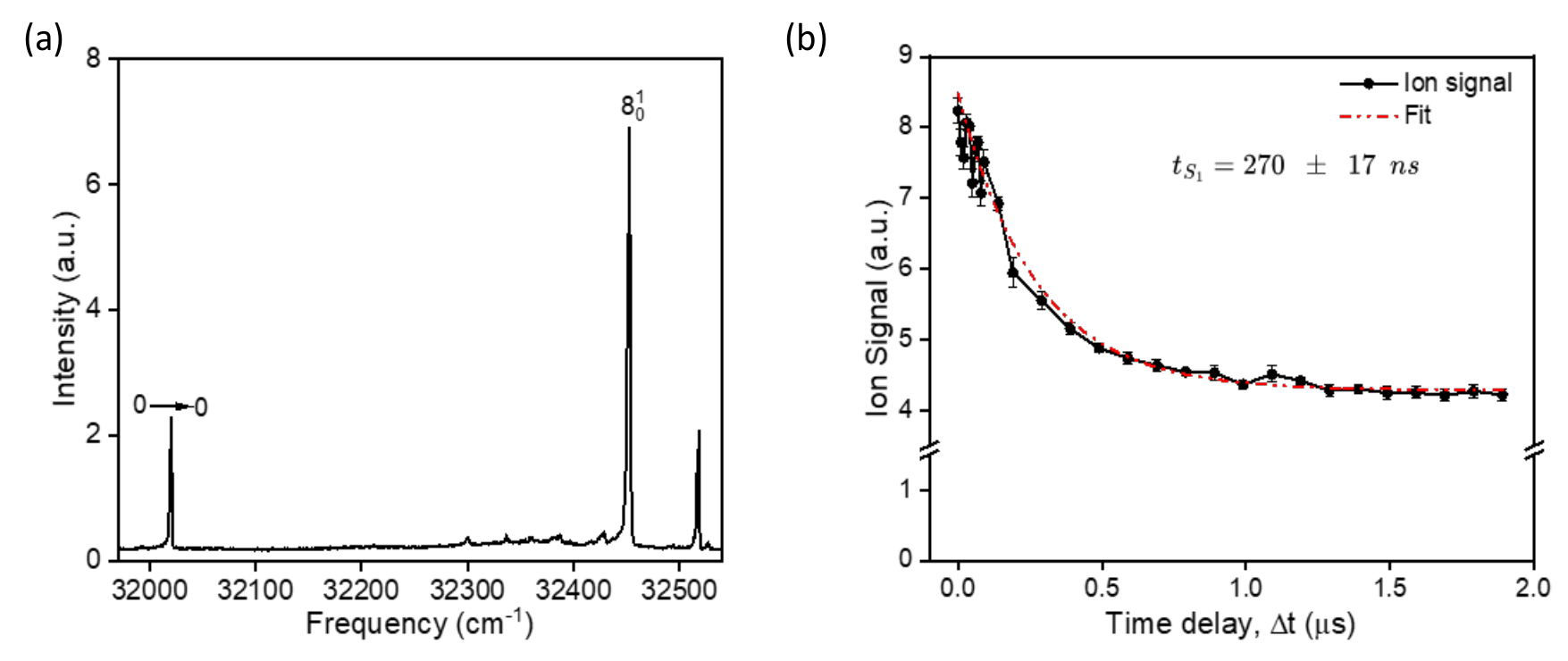}
    \caption{(a) R2PI spectrum of the S$_{0}$ state showing the S$_{0}$ $\leftarrow$ S$_{1}$ 0-0 transition at 32020 cm$^{-1}$. (b) Ion signal of naphthalene as a function of the time delay between the excitation and ionization lasers.}
    \label{Figure:S1}
\end{figure}

\begin{figure}
    \centering
    \includegraphics[width=0.6\linewidth]{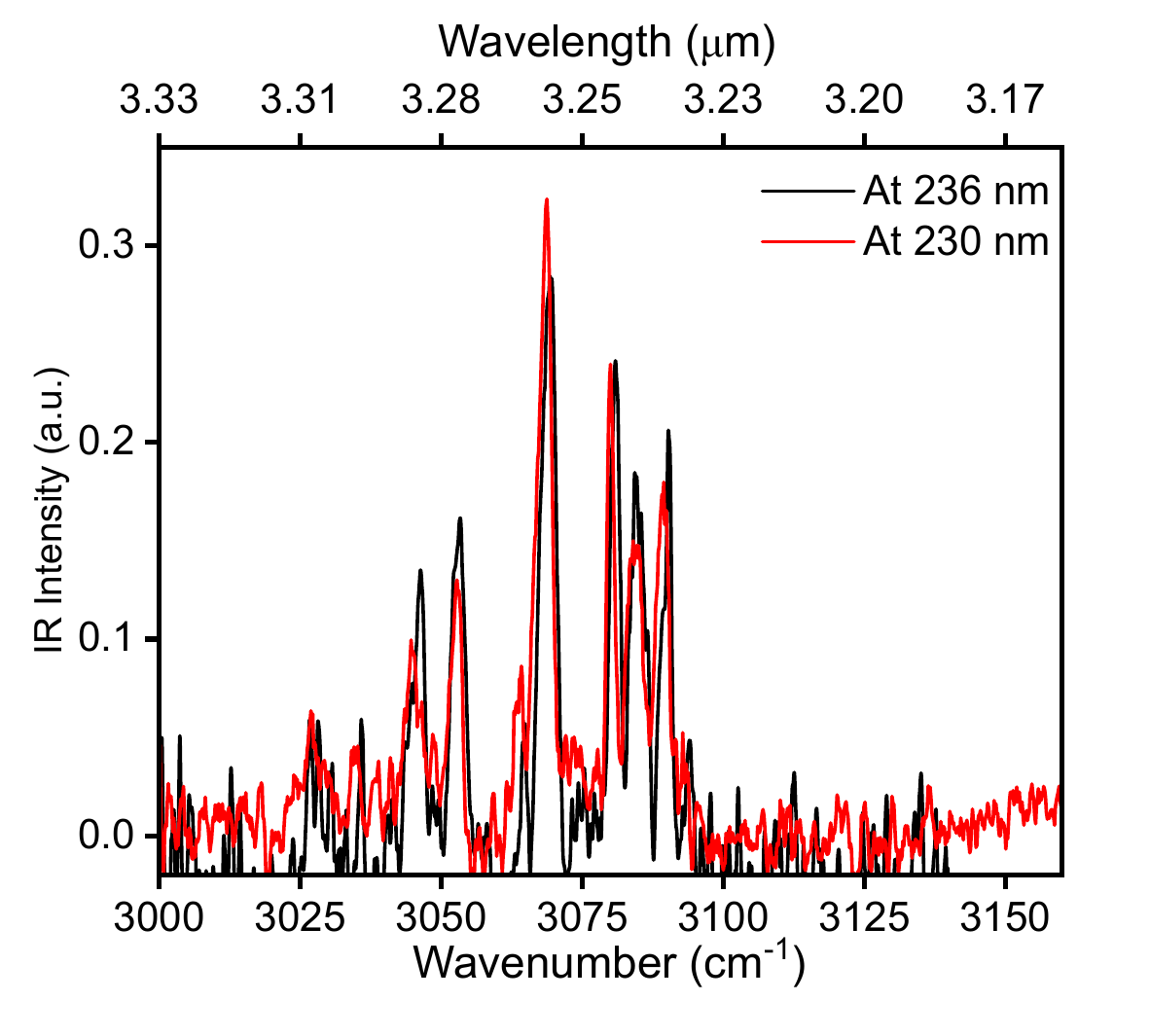}
    \caption{Experimental infrared absorption spectra of T$_{1}$ in the 3.33-3.16 $\mu$m (3000-3160 cm$^{-1}$) region obtained after excitation at 32020 cm$^{-1}$ and ionization at 236 nm (black) or 230 nm (red).}
    \label{Figure:S2}
\end{figure}

\begin{figure}
    \centering
    \includegraphics[width=0.9\linewidth]{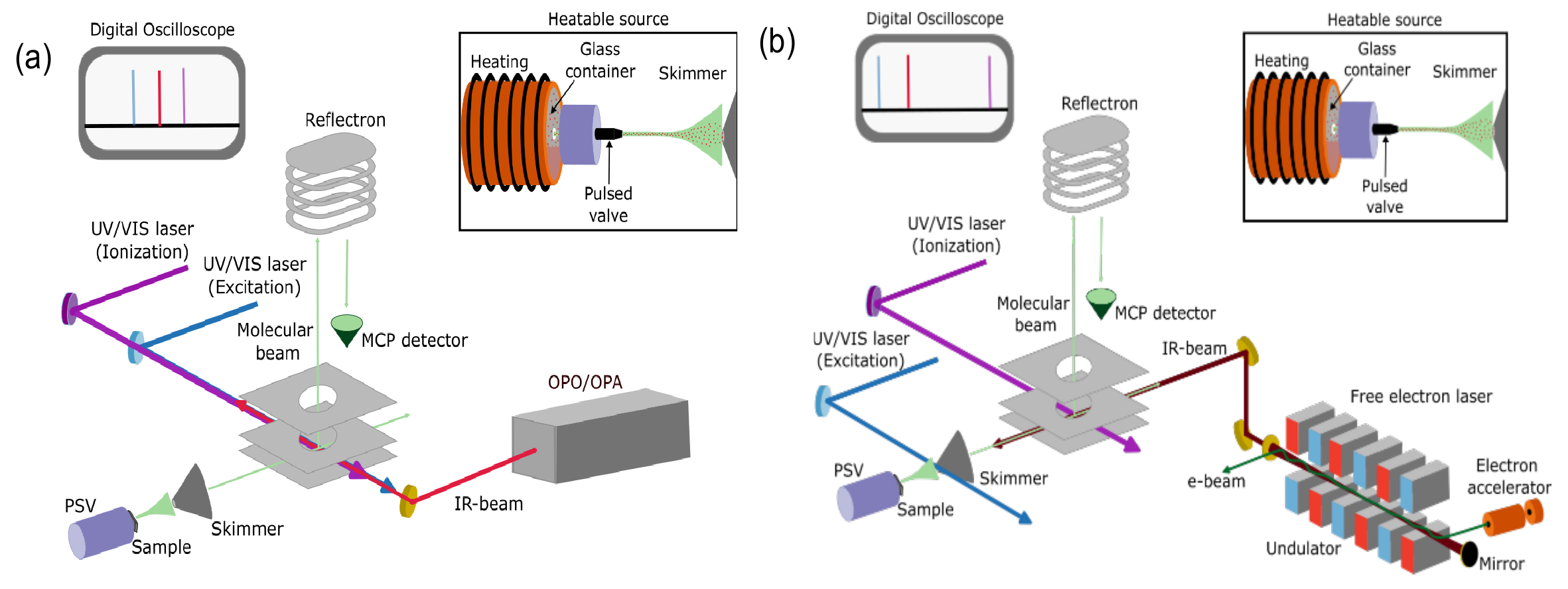}
    \caption{Schematic overview of the molecular beam setup incorporating two infrared laser sources: (a) OPO/OPA (3000–3140 cm$^{-1}$) and (b) FELIX (350–1700 cm$^{-1}$), covering different wavelength ranges. The setup consists of a heatable discharge source, skimmer, time-of-flight mass spectrometer, two UV lasers (excitation and ionization), and a digital oscilloscope. The oscilloscope is used to set the appropriate time delay between the excitation and ionization lasers and the IR sources. For the OPO/OPA laser, the IR beam is aligned perpendicular to the molecular beam, whereas for FELIX, the IR beam is aligned collinearly from the opposite end of the molecular beam. Adapted with permission from ref \cite{rijs2015ir} Copyright 2014 Springer Nature.}
    \label{Figure:S3}
\end{figure}

\begin{figure}
    \centering
    \includegraphics[width=0.6\linewidth]{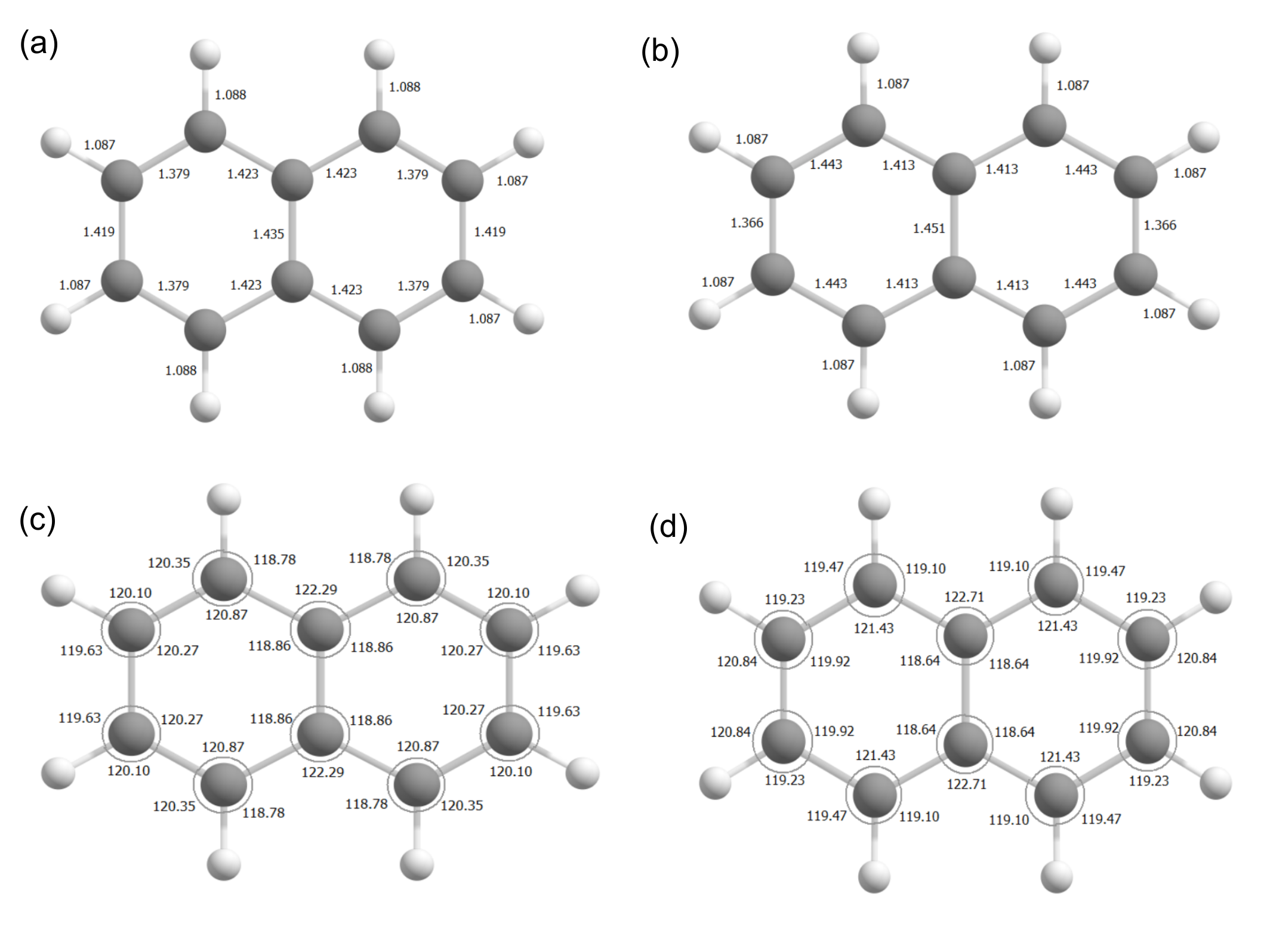}
    \caption{Structural parameters for (a, c) the S$_{0}$ and (b, d) the T$_{1}$ state of naphthalene. Figures (a) and (b) give the bond lengths, (c) and (d) the bond angles. The geometries have been optimized using the Gaussian 16 suite using the B3LYP functional with the N07D basis set.}
    \label{Figure:S4}
\end{figure}

\begin{figure}
    \centering
    \includegraphics[width=0.6\linewidth]{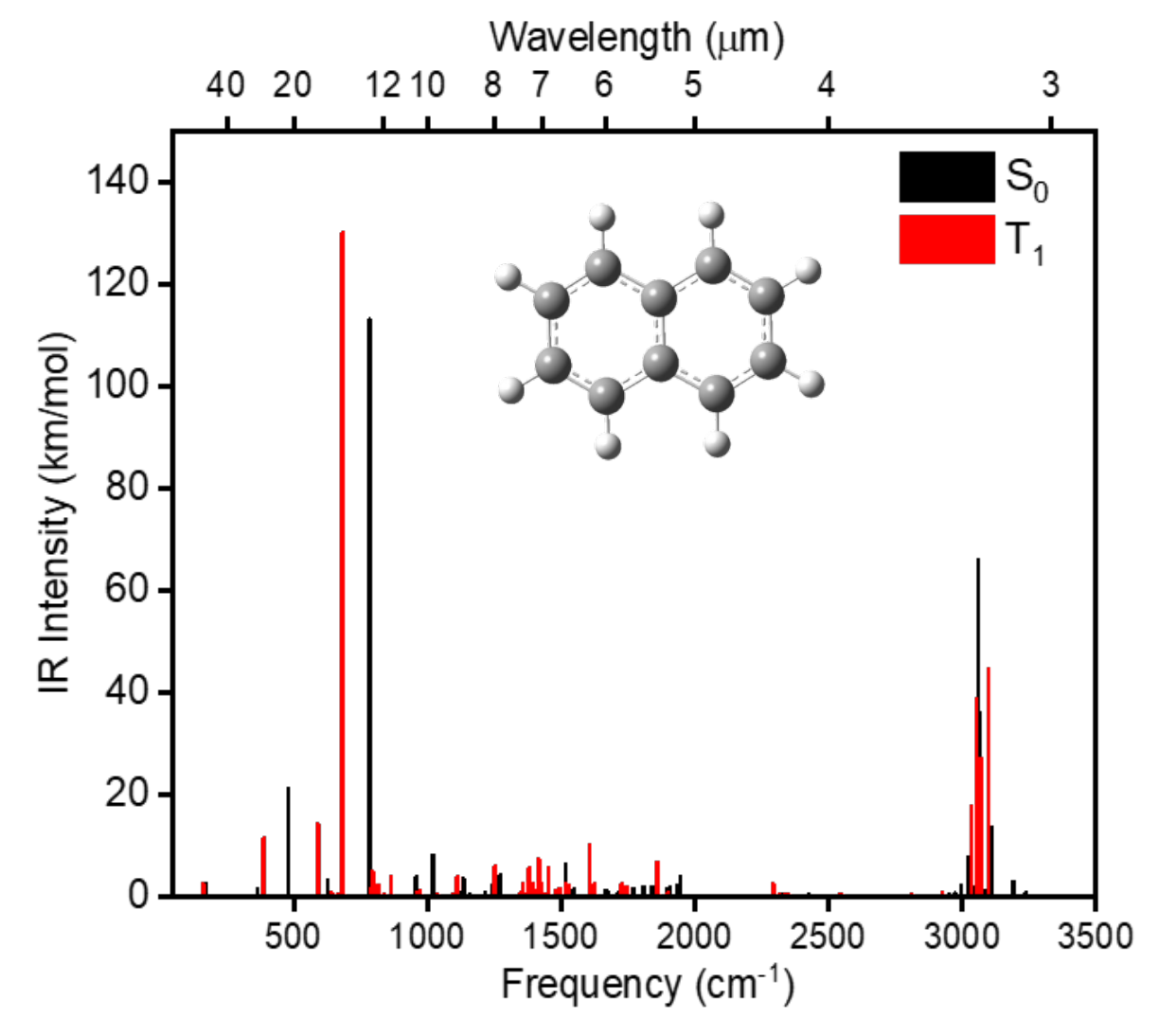}
    \caption{Computed anharmonic infrared spectra using GVPT2 at the B3LYP/N07D level for the S$_{0}$ (black) and T$_{1}$ states (red) of naphthalene.}
    \label{Figure:S5}
\end{figure}

\begin{figure}
    \centering
    \includegraphics[width=0.6\linewidth]{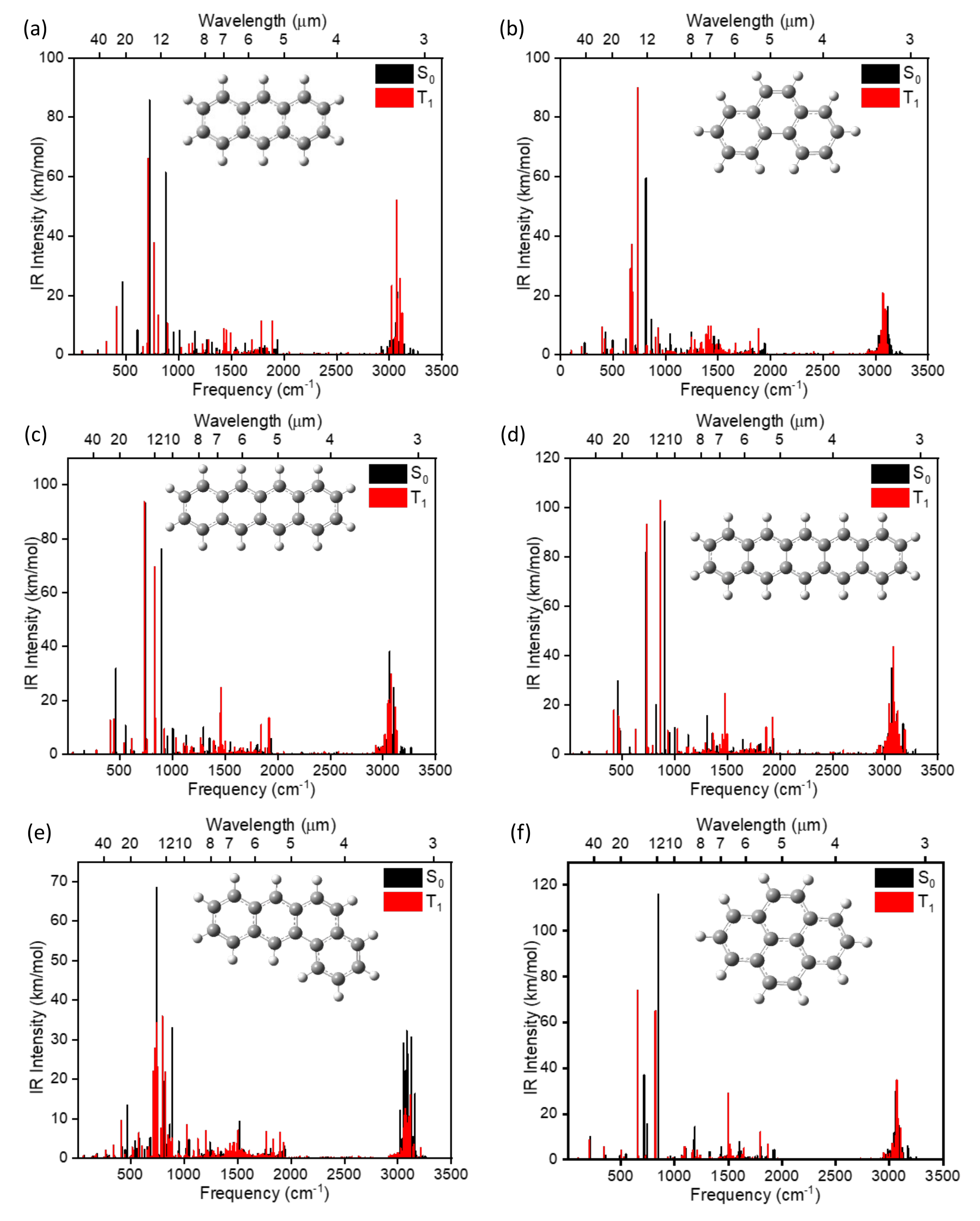}
    \caption{Computed anharmonic infrared spectra using ALPAHCAS (Reems \textit{et al.} in prep.) at the B3LYP/N07D level for the S$_{0}$ (black) and T$_{1}$ states (red) of (a) anthracene, (b) phenanthrene, (c) tetracene, (d) pentacene, (e) benz[a]anthracene, and (f) pyrene.}
    \label{Figure:S6}
\end{figure}

\begin{figure}
    \centering
    \includegraphics[width=0.6\linewidth]{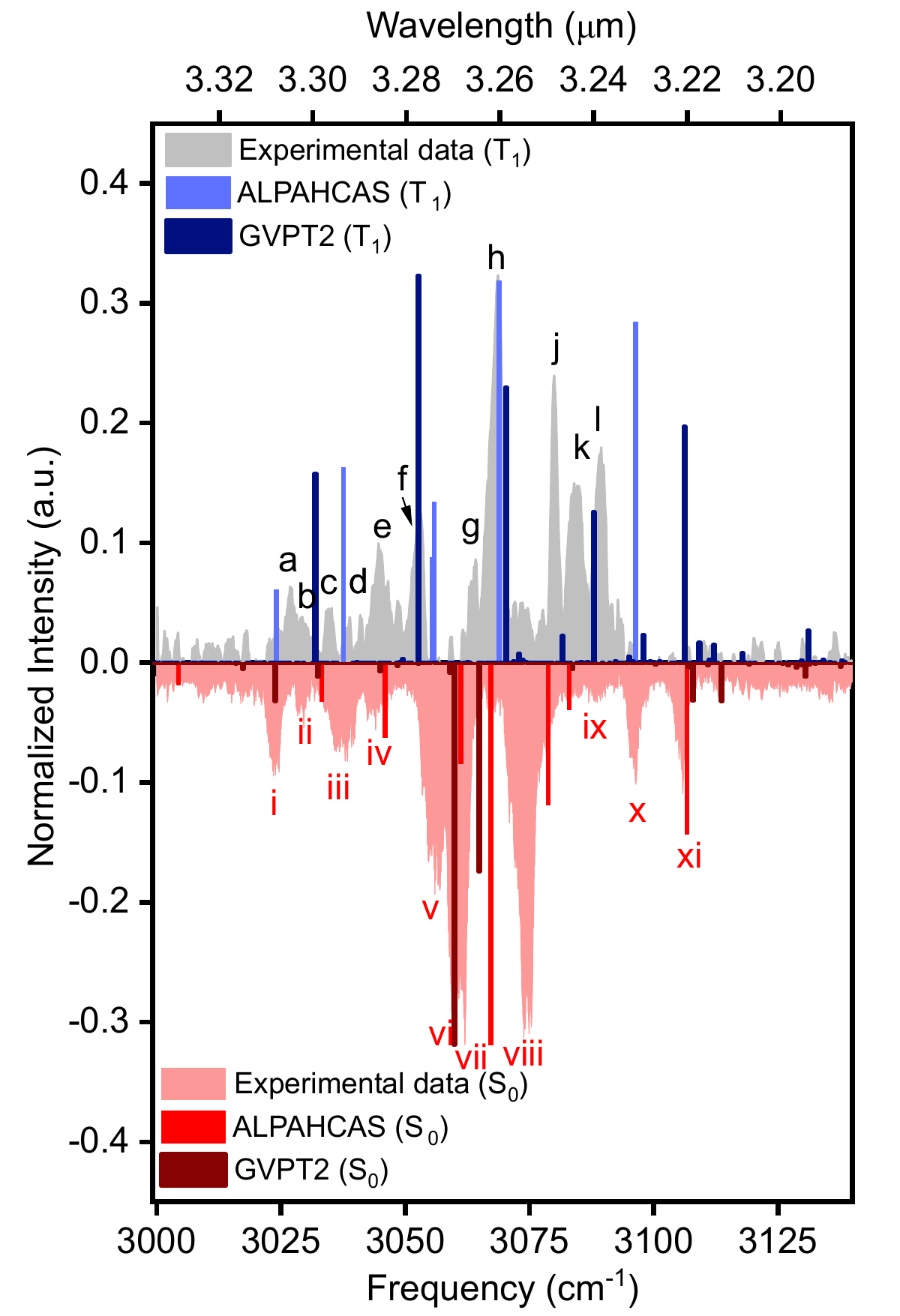}
    \caption{Experimental infrared absorption spectra of naphthalene in the T$_{1}$ (gray) and S$_{0}$ (red) states are shown in the 3.33-3.18 $\mu$m (3000-3140 cm$^{-1}$) region. The measured spectra are compared with anharmonic infrared spectra calculated using GVPT2 at the B3LYP/N07D level (dark blue and maroon stick spectra for T$_{1}$ and S$_{0}$, respectively) and with ALPAHCAS calculations (Reems et al., in prep.) (blue and red stick spectra for T$_{1}$ and S$_{0}$, respectively).}
    \label{Figure:S7}
\end{figure}

\begin{table} 
\textbf{Table A1.Naphthalene possesses D$_{2h}$ symmetry and comprises 48 fundamental vibrational modes. The vibrational frequencies, denoted $\nu_{1}$ to $\nu_{48}$, correspond to the fundamental modes ordered in descending frequency of naphthalene. The modes in the 3 $\mu$m region are compared with anharmonic vibrational frequencies obtained from GVPT2 (generalized vibrational second-order perturbation theory) calculations at the DFT level using Gaussian 16, as well as results from the ALPAHCAS code (Reems et al., in preparation). The entries under ID refer to the labeling of bands as given in Figure A7. (a) S$_0$ state - Relative intensities are normalized to the band at 3062 cm$^{-1}$ for comparison. \cite{mackie2015anharmonic}} \\
% ===================== (a) =====================
\begin{tabular}{l c c ccc ccc c}
\toprule
& Exp. & rel. I & \multicolumn{3}{c}{GVPT2} & \multicolumn{3}{c}{ALPAHCAS} & Symmetry \\
\cmidrule{4-6} \cmidrule{7-9}
ID & (cm$^{-1}$) &  & anharm & rel. I & Modes & anharm & rel. I & Modes & \\
\midrule
i  & 3023.9  & 0.28 & 3023.9 & 0.10 & $\nu_{7}$ & - & - & - & b$_{3g}$ \\
ii  & 3028.8 & 0.12 & 3032.4 & 0.04 & $\nu_{46} + 2\nu_{15}$ & 3033.2 & 0.10 & \begin{tabular}[t]{@{}c@{}}
 $\nu_{7}$ \\ 
 $\nu_{10} + \nu_{13}$ \\
 $\nu_{9} + \nu_{17}$
\end{tabular} & b$_{1u}$ \\
iii  & 3037.6 & 0.22 & - & - & - & - & - & - & - \\
iv  & 3044.1 & 0.16 & - & - & - & 3045.9 & 0.20 & \begin{tabular}[t]{@{}c@{}}
 $\nu_{9} + \nu_{15}$ \\ 
 $\nu_{10} + \nu_{14}$ \\
 $\nu_{2}$ \\ $\nu_{6}$
\end{tabular} & b$_{2u}$ \\
v  & 3056.2 & 0.59 & - & - & - & - & - & - & - \\
vi  & 3059.2  & 0.84 & 3059.9 & 1 & $\nu_{3}$ & - & - & - & b$_{1u}$ \\
vii  & 3062.0 & 1 & 3064.9 & 0.54 & $\nu_{2}$ & 3061.2 & 0.26 & \begin{tabular}[t]{@{}c@{}}
 $\nu_{6}$ \\ $\nu_{2}$ \\ 
 $\nu_{10} + \nu_{14}$ \\
 $\nu_{9}$ \\ $\nu_{15}$
\end{tabular} & b$_{2u}$ \\
viii  & 3074.2 & 0.94 & - & - & - & 3067.3 & 1 & \begin{tabular}[t]{@{}c@{}}
 $\nu_{3}$ \\ $\nu_{9} + \nu_{17}$ \\ 
 $\nu_{12} + \nu_{14}$ 
\end{tabular} & b$_{1u}$ \\
ix  & 3088.5  & 0.09 & - & - & - & 3082.9 & 0.12 &  \begin{tabular}[t]{@{}c@{}}
 $\nu_{10} + \nu_{13}$ \\ $\nu_{7}$ \\ $\nu_{3}$ \\ 
 $\nu_{9} + \nu_{12}$ 
\end{tabular} & b$_{1u}$ \\
x  & 3096.7 & 0.25  & - & - & - & - & - & - & - \\
xi  & 3105.9 & 0.34 & 3107.9 & 0.10 & $\nu_{11} +\nu_{12}$ & 3106.7 & 0.44 & 
\begin{tabular}[t]{@{}c@{}}
 $\nu_{11} + \nu_{12}$ \\ $\nu_{2}$ \\ $\nu_{6}$ \\ 
 $\nu_{10} + \nu_{14}$ 
\end{tabular} & b$_{2u}$ \\
\end{tabular}
\end{table}

\begin{table} 
\textbf{Table A1.Naphthalene possesses D$_{2h}$ symmetry and comprises 48 fundamental vibrational modes. The vibrational frequencies, denoted $\nu_{1}$ to $\nu_{48}$, correspond to the fundamental modes ordered in descending frequency of naphthalene. The modes in the 3 $\mu$m region are compared with anharmonic vibrational frequencies obtained from GVPT2 (generalized vibrational second-order perturbation theory) calculations at the DFT level using Gaussian 16, as well as results from the ALPAHCAS code (Reems et al., in preparation). The entries under ID refer to the labeling of bands as given in Figure A7. (b) T$_1$ state - Relative intensities are normalized to the band at 3069 cm$^{-1}$ for comparison.} \\
% ===================== (b) =====================
\begin{tabular}{l c c ccc ccc c}
\toprule
& Exp. & rel. I & \multicolumn{3}{c}{GVPT2} & \multicolumn{3}{c}{ALPAHCAS} & Symmetry \\
\cmidrule{4-6} \cmidrule{7-9}
ID & (cm$^{-1}$) &  & anharm & rel. I & Modes & anharm & rel. I & Modes & \\
\midrule
a  & 3027.0 & 0.19 & - & - & - & 3024.0 & 0.19 & $\nu_{9} +\nu{11}$ & b$_{2u}$ \\
b  & 3029.3 & 0.12 & - & - & - & - & - & - & - \\
c  & 3034.4 & 0.14 & 3031.9 & 0.90 & $\nu_{9} + \nu_{10}$ & 3037.6 & 0.51 & $\nu_{9} + \nu_{10}$ & b$_{2u}$ \\
d  & 3044.6 & 0.31 & - & - & - & - & - & - & - \\
e  & 3046.4 & 0.19 & - & - & - & 3055.4 & 0.27 & $\nu_{7}$ & b$_{2u}$ \\
f & 3052.8  & 0.41 & 3052.7 & 1.01 & $\nu_{6}$ & 3055.8 & 0.42 & $\nu_{6}$ & b$_{1u}$ \\
g  & 3062.9 & 0.20 & - & - & - & - & - & - & - \\
h  & 3064.2 & 0.27 & - & - & - & - & - & - & - \\
i & 3068.5  & 1.00 & 3070.3 & 1 & $\nu_{3}$ & 3068.8 & 1 & $\nu_{3}$ & b$_{1u}$ \\
j & 3080.0 & 0.75  & 3081.7 & 0.10 & $\nu_{9} + \nu_{24} + \nu_{42}$ & - & - & - & b$_{2u}$ \\
k  & 3084.0 & 0.47  & 3087.8 & 0.55 & $\nu_{9} + \nu_{32} + \nu_{35}$ & - & - & - & b$_{2u}$ \\
l  & 3089.1 & 0.54  & 3106.2 & 0.86 & $\nu_{2}$ & 3096.4 & 0.89 & $\nu_{2}$ & b$_{2u}$ \\
\end{tabular}
\end{table}

\begin{figure}
    \centering
    \includegraphics[width=0.8\linewidth]{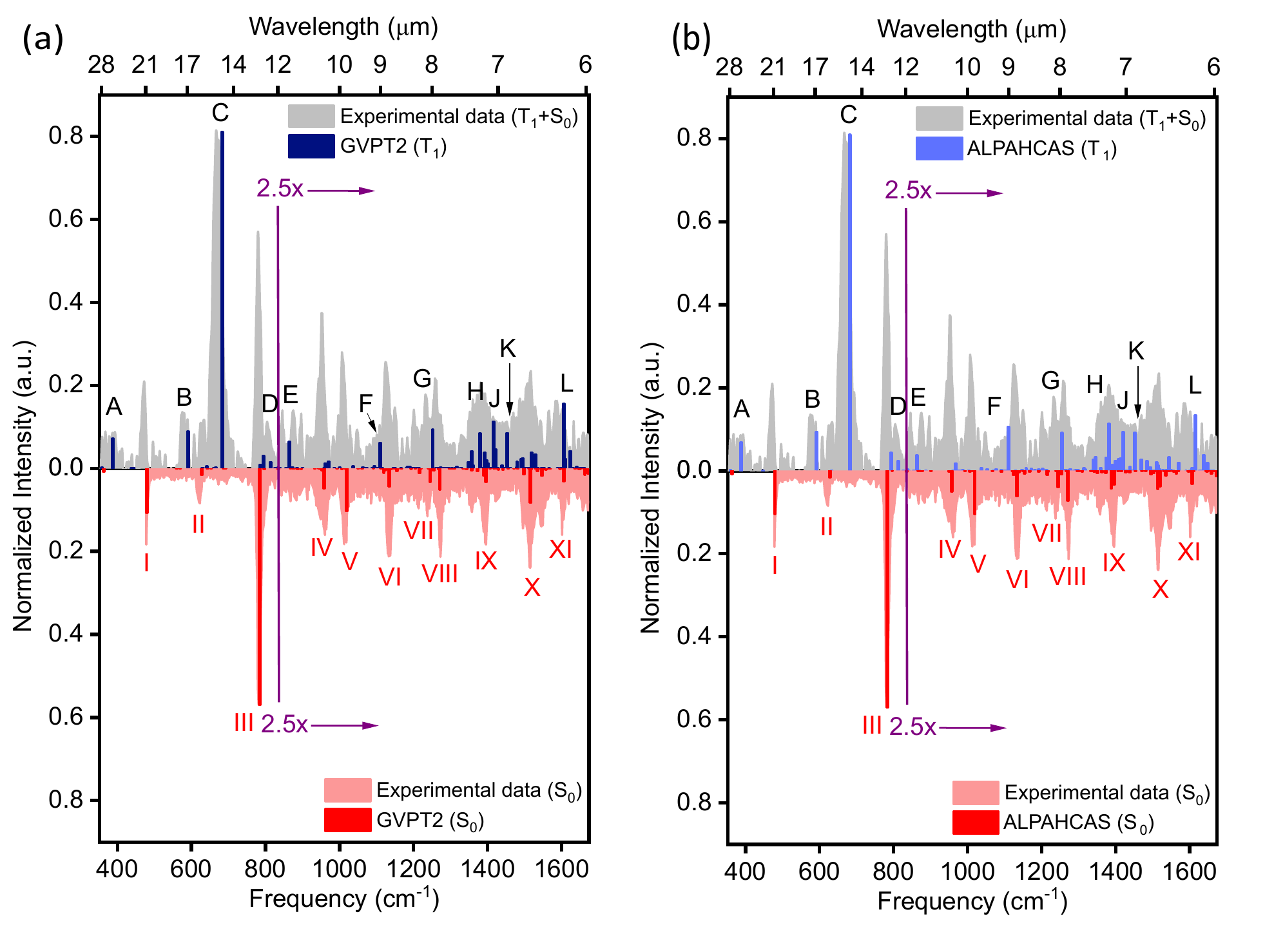}
    \caption{Experimental infrared absorption spectra of naphthalene in the T$_{1}$ (gray) and S$_{0}$ (red) states are shown in the 28.6-5.88 $\mu$m (350-1700 cm$^{-1}$) region. In panel (a), the measured spectra are compared with anharmonic infrared spectra calculated using GVPT2 at the B3LYP/N07D level (dark blue and maroon stick spectra for T$_{1}$ and S$_{0}$, respectively) and, in panel (b) they are compared with ALPAHCAS calculations (Reems et al., in prep.) (blue and red stick spectra for T$_{1}$ and S$_{0}$, respectively). For clarity, the relative intensities of the bands in the 830-1674 cm$^{-1}$ region are scaled by a factor of 2.5.}
    \label{Figure:S8}
\end{figure}

\begin{table}
\textbf{Table A2. Naphthalene possesses D$_{2h}$ symmetry and comprises 48 fundamental vibrational modes. The vibrational frequencies, denoted $\nu_{1}$ to $\nu_{48}$, correspond to the fundamental modes ordered in descending frequency of naphthalene. The modes in the fingerprint region are compared with anharmonic vibrational frequencies obtained from GVPT2 (generalised vibrational second-order perturbation theory) calculations at the DFT level using Gaussian 16, as well as results from the ALPAHCAS code (Reems et al., in preparation). The entries under ID refer to the labeling of bands as given in Figure A8. (a) S$_0$ state - Relative intensities are normalized to the band at 782 cm$^{-1}$ for comparison.\cite{lemmens2019anharmonicity}.} \\
% ===================== (a) =====================
\begin{tabular}{l c c ccc ccc c}
\toprule
& Exp. & rel. I & \multicolumn{3}{c}{GVPT2} & \multicolumn{3}{c}{ALPAHCAS} & Symmetry \\
\cmidrule{4-6} \cmidrule{7-9}
ID & (cm$^{-1}$) &  & anharm & rel. I & Modes & anharm & rel. I & Modes & \\
\midrule
I  & 478  & 0.32 & 481 & 0.30 & $\nu_{43}$ & 479 & 0.18 & $\nu_{43}$ & b$_{3u}$ \\
II  & 621 & 0.41 & 627 & 0.03 & $\nu_{39}$ & 627 & 0.03 & $\nu_{39}$ & b$_{2u}$ \\
III  & 782 & 1 & 784 & 1 & $\nu_{35}$ & 783 & 1 & $\nu_{35}$ & b$_{3u}$ \\
IV  & 960 & 0.11 & 958 & 0.03 & $\nu_{29}$ & 958 & 0.03 & $\nu_{29}$ & b$_{3u}$ \\
V  & 1019 & 0.12 & 1019 & 0.05 & $\nu_{26}$ & 1019 & 0.07 &  
\begin{tabular}[t]{@{}c@{}}
$\nu_{26}$ \\
$\nu_{40} + \nu_{41}$
\end{tabular} & b$_{2u}$ \\
VI  & 1133  & 0.15 & 1134 & 0.02 & $\nu_{24}$ & 1133 & 0.04 &  
\begin{tabular}[t]{@{}c@{}}
$\nu_{24}$ \\
$\nu_{37} + \nu_{46}$ \\
$\nu_{39} + \nu_{42}$
\end{tabular} & b$_{1u}$ \\
VII  & 1238 & 0.08 & 1245 & 0.01 & $\nu_{43} + \nu_{36}$ & 1244 & 0.03 & \begin{tabular}[t]{@{}c@{}}
$\nu_{36} + \nu_{43}$ \\
$\nu_{18}$
\end{tabular} & b$_{1u}$ \\
VIII  & 1273 & 0.15 & 1271 & 0.02 & $\nu_{18}$ & 1270 & 0.05 & 
\begin{tabular}[t]{@{}c@{}}
$\nu_{18}$ \\
$\nu_{36} + \nu_{43}$ 
\end {tabular} & b$_{1u}$ \\
IX  & 1394  & 0.12 & 1391 & 0.007 & $\nu_{15}$ & 1388 & 0.03 &  
\begin{tabular}[t]{@{}c@{}}
$\nu_{15}$ \\
$\nu_{25} + \nu_{46}$
\end{tabular} & b$_{1u}$ \\
X  & 1514 & 0.17  & 1516 & 0.03 & $\nu_{12}$ & 1513 & 0.03 & 
\begin{tabular}[t]{@{}c@{}}
$\nu_{12}$ \\
$\nu_{23} + \nu_{46}$
\end{tabular} & b$_{2u}$ \\
XI  & 1601 & 0.11 & 1606 & 0.01 & $\nu_{10}$ & 1606 & 0.02 & 
\begin{tabular}[t]{@{}c@{}}
$\nu_{10}$ \\
$\nu_{31} + \nu_{39}$
\end{tabular} & b$_{1u}$ \\
\end{tabular}
\end{table}

\begin{table}
\textbf{Table A2. Naphthalene possesses D$_{2h}$ symmetry and comprises 48 fundamental vibrational modes. The vibrational frequencies, denoted $\nu_{1}$ to $\nu_{48}$, correspond to the fundamental modes ordered in descending frequency of naphthalene. The modes in the fingerprint region are compared with anharmonic vibrational frequencies obtained from GVPT2 (generalised vibrational second-order perturbation theory) calculations at the DFT level using Gaussian 16, as well as results from the ALPAHCAS code (Reems et al., in preparation). The entries under ID refer to the labeling of bands as given in Figure A8. (a) S$_0$ state - Relative intensities are normalized to the band at 782 cm$^{-1}$ for comparison.\cite{lemmens2019anharmonicity}. (b) T$_1$ state - Relative intensities are normalized to the band at 667 cm$^{-1}$ for comparison.} \\
% ===================== (b) =====================
\begin{tabular}{l c c ccc ccc c}
\toprule
& Exp. & rel. I & \multicolumn{3}{c}{GVPT2} & \multicolumn{3}{c}{ALPAHCAS} & Symmetry \\
\cmidrule{4-6} \cmidrule{7-9}
ID & (cm$^{-1}$) &  & anharm & rel. I & Modes & anharm & rel. I & Modes & \\
\midrule
A  & 386 & 0.10 & 388 & 0.09 & $\nu_{44}$ & 388 & 0.09 & $\nu_{44}$ & b$_{3u}$ \\
B  & 575 & 0.17 & 590 & 0.11 & $\nu_{39}$ & 592 & 0.11 & $\nu_{39}$ & b$_{2u}$ \\
C  & 667 & 1 & 683 & 1 & $\nu_{37}$ & 682 & 1 & $\nu_{37}$ & b$_{3u}$ \\
D  & 804 & 0.14 & 795 & 0.04 & $\nu_{33}$ & 793 & 0.05 & $\nu_{33}$ & b$_{1u}$ \\
E  & 875 & 0.07 & 865 & 0.03 & $\nu_{29}$ & 863 & 0.02 & $\nu_{29}$ & b$_{2u}$ \\
F  & 1107  & 0.03 & 1110 & 0.03 & $\nu_{22}$ & 1110 & 0.05 & $\nu_{22}$ & b$_{2u}$ \\
G  & 1234 & 0.08 & 1252 & 0.05 & $\nu_{18}$ & 1255 & 0.04 & \begin{tabular}[t]{@{}c@{}}
$\nu_{18}$ \\
$\nu_{29} + \nu_{42}$\\
$\nu_{30} +\nu_{43}$
\end{tabular}
& b$_{1u}$ \\
H  & 1372 & 0.09 & 1381 & 0.04 & $\nu_{15}$ & 1383 & 0.06 & \begin{tabular}[t]{@{}c@{}}
$\nu_{15}$ \\
$\nu_{23} + \nu_{45}$ \\
$\nu_{36} +\nu_{37}$
\end{tabular} & b$_{1u}$ \\
J & 1419  & 0.06 & 1416 & 0.06 & $\nu_{13}$ & 1421 & 0.04 & \begin{tabular}[t]{@{}c@{}}
$\nu_{13}$ \\
$\nu_{23} + \nu_{45}$ \\
$\nu_{29 + \nu_{42}}$
\end{tabular}& b$_{1u}$ \\
K  & 1464 & 0.05  & 1454 & 0.04 & $\nu_{11}$ & 1452 & 0.04 & 
\begin{tabular}[t]{@{}c@{}}
$\nu_{11}$ \\
$\nu_{10}$ \\
$\nu_{31} + \nu_{37}$
\end{tabular}& b$_{2u}$ \\
L  & 1584 & 0.08  & 1606 & 0.08 & $\nu_{30} + \nu_{31}$ & 1616 & 0.06 & 
\begin{tabular}[t]{@{}c@{}}
$\nu_{30} + \nu_{31}$ \\
$\nu_{32} + \nu_{35}$ \\
$\nu_{29} + \nu_{34}$
\end{tabular} & b$_{2u}$ \\
\end{tabular}
%\begin{tablenotes}
%\small
%\item Note: This is an example of a table footnote.
%\item[1] Example for a first table footnote.
%\item[2] Example for a second table footnote.
%\end{tablenotes}
\end{table}

\begin{figure}
    \centering
    \includegraphics[width=0.6\linewidth]{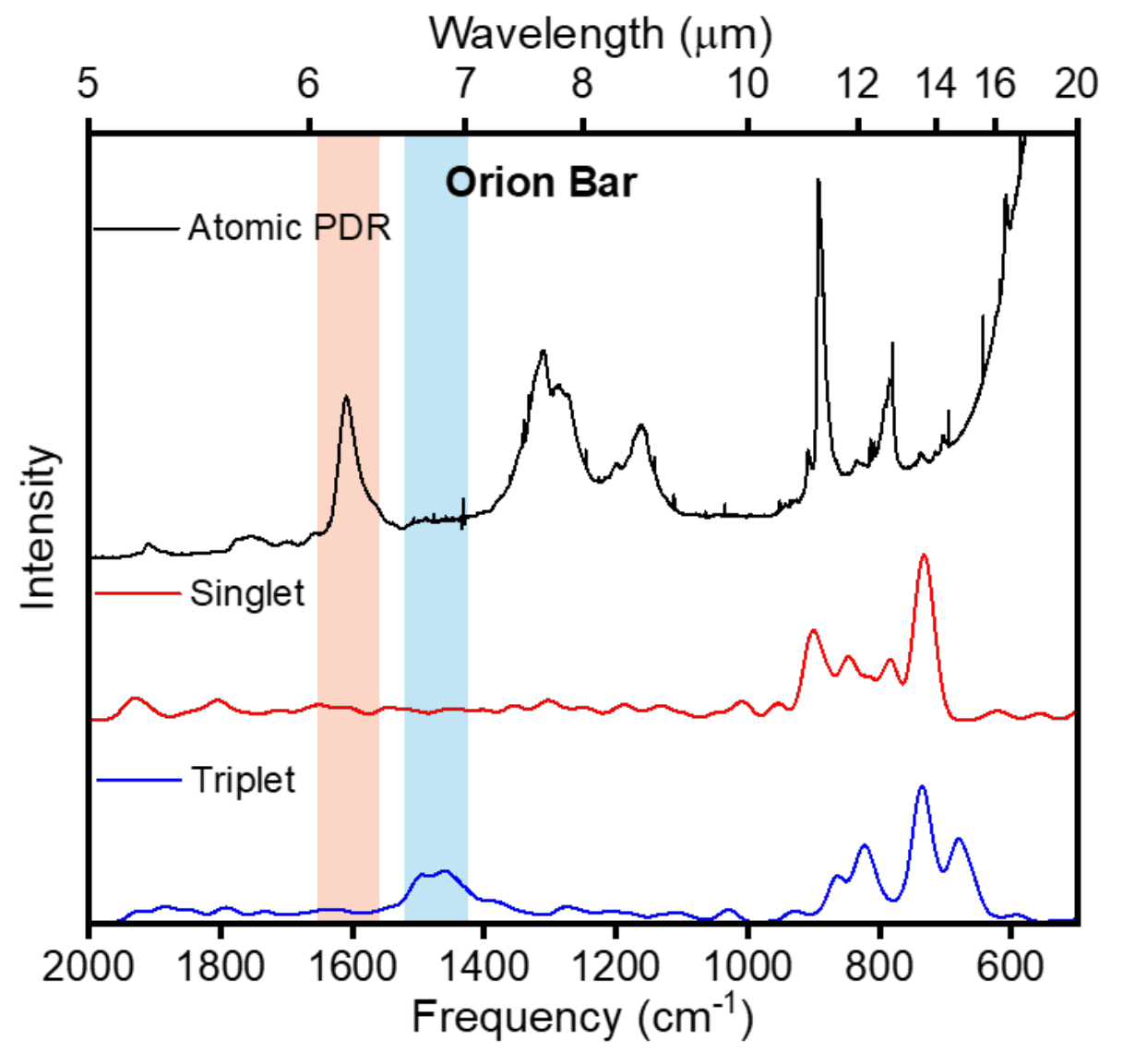}
    \caption{Comparison of the JWST/MIRI spectrum of the Atomic PDR region in the Orion Bar (black) with averaged theoretical vibrational spectra of seven neutral PAHs (naphthalene, anthracene, phenanthrene, tetracene, benz[a]anthracene, pentacene, and pyrene). The calculated spectra correspond to the singlet (red) and triplet (blue) electronic states. Anharmonic vibrational frequencies and intensities were obtained from ALPAHCAS calculations (Reems \textit{et al.} in prep.). Stick spectra were convoluted with Gaussian profiles (FWHM = 30 cm$^{-1}$) and averaged prior to comparison with the observed spectrum. In the observational data, the 6.2 $\mu$m band is highlighted in light red, while the 6.8 $\mu$m simulated feature is shown in light blue.}
    \label{Figure:S9}
\end{figure}

\bibliography{sn-bibliography}

\end{document}